\documentclass[conference]{IEEEtran}
\IEEEoverridecommandlockouts
\usepackage[bottom=1.06in,top=0.75in,left=0.68in,right=0.68in]{geometry}
\usepackage{amsmath}
\usepackage{amsfonts}
\usepackage{amssymb}
\usepackage[ruled,vlined]{algorithm2e}
\usepackage{caption}
\usepackage{mathtools}  
\usepackage{amsthm}
\usepackage[]{algorithm2e}
\usepackage{booktabs}
\usepackage{subfigure}
\usepackage{tabulary}
\usepackage{footnote}
\usepackage[export]{adjustbox}
\usepackage{booktabs}
\usepackage[justification=centering]{caption}
\usepackage{comment}

\usepackage{cases}
\usepackage{graphicx}
\usepackage{cite}
\usepackage{multicol}
\usepackage{xcolor}
\usepackage{graphicx}
\graphicspath{{./}{figures/}}
\newcommand\numeq[1]%
  {\stackrel{\scriptscriptstyle(\mkern-1.5mu#1\mkern-1.5mu)}{=}}
\usepackage{stfloats}% <-- added
\usepackage{cuted}
\usepackage{lipsum}  

\usepackage{subfigure}
\usepackage{dsfont}
\usepackage{amsmath,amssymb}
\newcommand{\tabincell}[2]{\begin{tabular}{@{}#1@{}}#2\end{tabular}}

\title{Resource Allocation for Vehicle Platooning in 5G NR-V2X via Deep Reinforcement Learning}

\author{\IEEEauthorblockN{Liu Cao$^*$$^,$$ ^\dagger$, Hao Yin$^*$$^,$$ ^\dagger$}
\IEEEauthorblockA{{$^*$Department of Electrical and Computer Engineering, University of Washington, Seattle, USA} \\{$^\dagger$
Department of Applied Mathematics, University of Washington, Seattle, USA}\\
Email: \{liucao, haoyin\}@uw.edu } 
}
\begin{document}

\maketitle
\thispagestyle{empty}
\begin{abstract}
Vehicle platooning, one of the advanced services supported by 5G New Radio V2X (NR-V2X), improves traffic efficiency in the connected intelligent transportation systems (C-ITSs). However, the packet collision probability of platoon communication, especially in the out-of-coverage area, is significantly impacted by the random selection algorithm employed in the current resource allocation scheme. In this paper, we first analyze the collision probability via the random selection algorithm based on the current standard. Subsequently, we investigate the deep reinforcement learning (DRL) algorithm that decreases the collision probability by letting the agent (platoon leader) learn from the communication environment. Monte Carlo simulation is utilized to verify the results obtained in the analytical model and to compare the results between the two discussed algorithms. Numerical results show that the proposed DRL algorithm outperforms the random selection algorithm in terms of different vehicle density, which at least lowering the collision probability by 73\% and 45\% in low and high vehicle density respectively.

\end{abstract}


\section{Introduction}
\label{sec:intro}
Vehicle-to-Everything (V2X) communications, which support traffic safety, traffic efficiency and advanced V2X services, are paving the way towards the connected intelligent transportation systems (C-ITSs). Cellular V2X (C-V2X), initially defined as LTE-V2X in 3GPP Release 14 \cite{3gpp.36.300}, is an example of a leading technology in V2X communications. Furthermore, New Radio V2X (NR-V2X) was designed in 3GPP Release 16 (5G phase) to support the advanced services that primarily focus on vehicle platooning, advanced driving, extended sensor and remote driving \cite{3gpp.22.886}. Different from LTE-V2X, which only provisions support for broadcast transmissions, NR-V2X allows multiple communication types to act simultaneously \cite{naik2019ieee}: A platoon leader (PL) in the platoon can communicate with its platoon members (PMs) using the groupcast mode, while utilizing the broadcast mode to transmit other periodic messages to the vehicles that are not part of the platoon. Note that in vehicle platooning, a PL coordinates the movement of a group of PMs by broadcasting Basic Safety Messages (BSMs) cyclically \cite{giambene2020analysis}. 

\begin{figure*}[t]
    \centering
    \includegraphics[width=.8\textwidth]{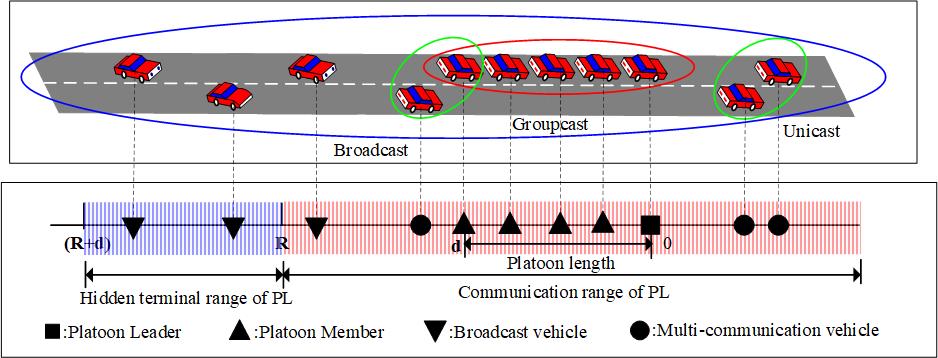}
    \caption{A multi-lane highway where multiple communication types coexist for vehicular communications.}
    \label{fig:VANET}
\end{figure*}

NR-V2X defines two sidelink (NR-PC5) modes: The NR-V2X sidelink Mode-1 signifies the mechanisms that allow direct vehicular communications within gNodeB coverage. The gNodeB in this mode schedules the sidelink resources for all vehicles. The NR-V2X sidelink Mode-2 supports direct vehicular communications in the out-of-coverage scenario \cite{3gpp.38.885}. In NR-V2X Mode-2, the vehicle selects the resources for V2X communication autonomously: vehicles perform sensing on a set of resources, i.e., resource pool, to avoid packet collisions for V2X message transmission\cite{ganesan20195g}. The general resource allocation scheme in the current standard for Mode 2 is to randomly select a few of the idle resources periodically. However, in practice, the randomness utilized in the general scheme does not solve some issues such as the hidden terminal effects and resource selection collision problem effectively, still incurring a high collision probability. Therefore, it is necessary to decrease the collision probability as the packet delivery of PL-to-PM demands high reliability in platoon communication.

A wide array of studies and new use cases have investigated the reliability enhancement for platoon communication. The authors in \cite{li2020resource,mei2018joint} studied and implemented some novel schemes to guarantee the reliability requirements of V2V communication and string stability of platoon based on LTE-V2V. The literature \cite{yu2018modeling ,cao2020VTC} provided the performance analysis of 802.11p-based communication from different perspectives. A platoon communication mode based on
D2D technology was proposed in \cite{wang2018resource} to share inter-vehicle control information
efficiently and timely. In \cite{wang2019platoon}, the authors developed a two-step sub-channel allocation strategy by which the base station and each platoon leader allocated the resources to improve the multi-platoon cooperations. In addition, it was shown in \cite{hegde2019enhanced} that a group scheduling mechanism enabled to overcome the cell border issues where the platoon members of a long platoon are served by different base station. As a result, most existing work focus on the platoon communication within base station (BS) coverage. Furthermore, to our best knowledge, there is not sufficient study on the resource allocation for platoon communication which takes adequate consideration of the current NR-V2X standard. 

Motivated by the aforementioned points, our work features the new characteristics of NR-V2X where multiple communication types coexist, concentrating on the resource allocation for platoon communication in the out-of-coverage scenario based on a simplified version of Mode 2. We analyze the random selection algorithm in the current standard as the baseline algorithm. Then we apply the deep reinforcement learning algorithm to decrease the collision probability by letting PL learn from the communication environment due to the potential benefits of learning from the environment \cite{Yin2019TCOM, Yin2020WNS3, DRLVTC}. In addition, a Monte Carlo simulator is employed to verify the results obtained by the analytical model and to compare the results between the two discussed algorithms. The major contributions of this paper are summarized as follows.
\begin{itemize}
\item We investigate the resource allocation for platoon communication from the MAC-layer perspective in the NR-V2X supported scenario where unicast, groupcast and broadcast communications coexist, and all of the vehicular communications operate without the assistance of BS.
\item We propose employing the deep reinforcement learning in resource allocation problem for platoon communication, which exhibits that PL can learn from interactions with the communication environment and figure out a clever policy of collaborating with PMs to improve the communication reliability based on the received feedback.
\end{itemize}

The rest of this paper is organized as follows: Section \ref{sec:sys} analyzes the collision probability with random selection algorithm. In section \ref{sec:DRL}, we propose utilizing the deep reinforcement learning algorithm to the resource allocation problem for platoon communication. Section \ref{sec:sim} verifies the analytical model of the random selection algorithm by the simulations, and then compares the results between the two discussed algorithms. Finally, section \ref{sec:con} draws the conclusions.

\section{System model}
\label{sec:sys}
\subsection{System Architecture}
We consider a finite length multi-lane highway where one platoon with length $d$ (km), broadcast vehicles and unicast vehicles are all considered to share the same resource pool for vehicular communications, as Fig. \ref{fig:VANET} illustrates. The 2-dimensional scenario can be mapped into a 1-dimensional finite line scenario where the location of each vehicle is assumed to be uniformly distributed in the line with density $\rho$ (vehicles/km). Meanwhile, we assume that the location of the vehicles remains the same, i.e., all vehicles are assumed to be static. Following the standard's guidelines\cite{3gpp.37.885}, we define the scenario, V2V-Highway where a pair of vehicles can be in one of following states:
\begin{itemize}
\item Line-of-Sight (LOS), if the vehicles are in the same street and the path is free from obstacles;
\item Vehicle Non-Line-of-Sight (NLOSv), if the vehicles are in the same street but the path is blocked by other vehicles.
\end{itemize} 
The link pathloss is implemented accurately following the 3GPP specifications \cite{3gpp.37.885}. The LOS/NLOSv pathloss is 
\begin{equation}\label{eq:PLNLOSv}
    PL_{\text{LOS/NLOSv}} = 32.4+20\cdot \text{log}_{10}D +20\cdot \text{log}_{10}f_c [dB],
\end{equation}
where $D$ is the distance in meters, and $f_c$ is the center frequency in GHz. Therefore, if the sensitivity-power-level at the receiver is set to $P_{sen}$ while the transmit power of each broadcast vehicle is $P_{trans}$\footnote{The standard \cite{3gpp.36.101} specifies that the maximum transmit power is 23dBm, and a sensitivity-power-level requirement at the receiver is -91dBm.}, the radius of communication range of each broadcast vehicle can be determined by
\begin{equation}\label{eq:range}
    \text{log}_{10} R = \frac{(P_{trans}-P_{sen}-92.4-20\cdot \text{log}_{10}f_c)}{20},
\end{equation}
where $R$ is in kilometers. Notice that the hidden terminal range of a platoon, from the side of PL, is also equal to the platoon length. The relation between the communication range of PMs and PL can be summarized as: 
\begin{equation}\label{eq:R}
    R_{PM_n}\subseteq \{R_{PM_m}\cup R_{PL}\},
\end{equation}
where $R_{PM_n}$ and $R_{PL}$ denote the communication range of PMs and PL. $n \in \{1,2,...,m\}$ is the PM index, and $m$ is the index of the last PM. As a result, in a 1-dimensional finite line scenario, the last PM suffers most from the hidden terminal effects among all PMs, determining the collision probability of the platoon communication. To ensure reliable group communication, PL allocates group internal identity to each of the PMs to help configure and distribute sidelink hybrid automatic
repeat request (SL-HARQ) feedback resources \cite{ganesan2020nr}: Each PM transmits either HARQ-Positive Acknowledgement (HARQ-ACK) or HARQ-Negative Acknowledge (HARQ-NACK) on a given set of dedicated feedback resources \cite{3gpp.38.885}.

In the NR-V2X system, the channel is divided into $t_s$ ms slot and $n_r$ sub-channels with a set of resource blocks (RBs), each comprised of a group of OFDMA tones. Consider the broadcast vehicles and PL regularly generate BSMs every $T_{tr}$ ms which is one transmission period. Each transmission period contains $N_r$ virtual resource blocks (VRBs), and each VRB can support one BSM transmission in a slot. Note that in each slot, the resources can support a fixed number of BSM transmissions, i.e., we can view the resources in one slot as $n_r$ VRBs. Accordingly, the total number of possible VRBs that one vehicle can select in each transmission period is given by
\begin{equation}\label{eq:N_r}
    N_r = \frac{T_{tr}n_r}{t_s}.
\end{equation}

Fig. \ref{fig:RB_selection} shows the VRB maps of resource allocation for platoon communication. In each transmission period, PL randomly selects one of the VRBs that were sensed as idle from the last transmission period, and then broadcasts the BSM to the PMs. The leading factors in the collision mainly include the resource selection collision and hidden terminal effects: The resource selection collision occurs when the broadcast vehicles in the communication range of PL select the same VRB as the PL while the hidden terminal effects 
generate when the broadcast vehicles in the hidden terminal range of PL select the same VRB as the PL.

The broadcast vehicles in the NR-V2X system adopt a distributed resource reservation algorithm-a sensing-based Semi-Persistent Scheduling (SPS) scheme\cite{3gpp.36.213,3gpp.36.321}. During the semi-persistent period, consisting of $T_s$ transmission periods, each vehicle keeps its current VRB selection, even if a collision occurs, since there is no ACK/NACK when broadcasting the BSMs. When a new semi-persistent period starts, each vehicle continues sticking to the VRB selection with a resource keeping probability $p$. Each time when a vehicle wants to change its VRB selection, it will randomly select one of the VRBs with lowest 20\% of transmission power, sensed from the last transmission period. If we only consider the MAC-layer performance, all vehicles only sense whether the VRB is occupied over the last transmission period, and are synchronized in terms of the semi-persistent period. Since vehicles can start broadcasting the packet right after it generates, allowing vehicles to select the closest idle VRB will reduce the average delay\cite{wang2018fixed}. In this paper, each broadcast vehicle selects the closest idle VRB with the probability $1-p$ when a new semi-persistent period starts.

\subsection{Analytical Model of Random Selection Algorithm}
In this section, we construct an analytical model of random selection algorithm for platoon communication. We first analyze the collision probability for the resource selection collision. Consider the case without the participation of all possible hidden terminals of PL. In other words, the current vehicular network is a fully connected network where PL can see all the vehicles only in its communication range. In one transmission period, the probability that $n$ out of $2R\rho -1$ vehicles, which is the average number of vehicles in the communication range of PL, decide to change their VRB selections is 
\begin{equation}\label{eq:P_r}
    P_r(n) = \begin{pmatrix} 2R\rho-1 \\ n \end{pmatrix} \left(\frac{1-p}{T_s}\right)^n\left(1-\frac{1-p}{T_s}\right)^{2R\rho -1-n}.
\end{equation}
When $n$ vehicles decide to change their VRB selections at the same time as PL, the probability that at least one of the $n$ vehicles collides with PL is 
\begin{equation}\label{eq:P_s}
    P_s(n) = 1- \left(\frac{N_r-N_a}{N_r}\right)^{n},
\end{equation}
where $N_a$ is the average number of VRBs which one vehicle must occupy in advance to collide with PL. $N_a$ is approximately equal to 1 in low vehicle density while being greater than 1 in high vehicle density, which can be written as
\begin{equation}\label{eq:N_a}
    N_a = \sum_{h=0}^{N_r -2}\left[1-\left(1-\frac{1}{N_r}\right)^{2R\rho-1}\right]^h,
\end{equation}
where $h$ is the number of VRBs between the VRB that one vehicle has occupied and the VRB that both PL and this vehicle will select. If $N_r$ is large enough, the approximate value of $N_a$ is given by
\begin{equation}\label{eq:N_a_approx}
    N_a \approx  \frac{1}{\left(1-\frac{1}{N_r}\right)^{2R\rho-1}}.
\end{equation}
Therefore, the collision probability for the resource selection collision is
\begin{equation}\label{eq:P_c^sr_1}
    P_{c}^{rs} = \sum_{n=1}^{2R\rho -1}P_r(n)P_s(n).
\end{equation}
We can substitute equation (\ref{eq:P_r}) and (\ref{eq:P_s}) into (\ref{eq:P_c^sr_1}) and get the following equation by using binomial theorem:
\begin{equation}\label{eq:P_c^sr}
    P_{c}^{rs} = 1- \left[ 1- \frac{(1-p)N_a}{T_s N_r}\right]^{2R\rho-1}.
\end{equation}

Now we focus on the collision with the hidden terminal effects. Since the broadcast vehicles in the hidden terminal range also use the same resource allocation scheme, each of them can see the VRBs occupied by other vehicles in its own communication range. Thus vehicles in the hidden terminal range of PL can see at most $R\rho$ VRBs occupied by other vehicles in the communication range of PL. Assume each broadcast vehicle in the hidden terminal range does not use the $R\rho$ VRBs and its VRB selection is uniformly distributed in the rest $N_r - R\rho$ VRBs. The probability that one broadcast vehicle in the hidden terminal range does not collide with PL is
\begin{equation}\label{eq:P_one^sr}
    P_{one}^{ht}= \frac{N_r-R\rho -1}{N_r-R\rho}.
\end{equation}
If one BSM from PL is successfully delivered to the PMs with the hidden terminal effects, two conditions must be satisfied: The resource selection collision does not occur as well as the broadcast vehicles in the hidden terminal range do not select the same VRB as PL. Note that the average number of hidden terminals for the platoon, from the side of PL, is $d\rho$, thus, the collision probability considering the hidden terminal effects can be expressed as
\begin{equation}\label{eq:P_c^ht}
    P_{c}^{ht}= 1-\left(1-P_c^{rs}\right)\left(P_{one}^{ht}\right)^{d\rho}.
\end{equation}

As illustrated before, the random selection algorithm also incurs a high collision probability. According to the characteristics of platoon communication, the PL, regarded as an agent, can explore the unknown communication environment based on the received feedback from the PMs. We next introduce the deep reinforcement learning to effectively decrease the collision probability, $P_{c}^{ht}$.

\section{Deep Reinforcement Learning}
\label{sec:DRL}
\subsection{Markov Decision Process}
\begin{figure}[t]
    \centering
    \includegraphics[width=.45\textwidth]{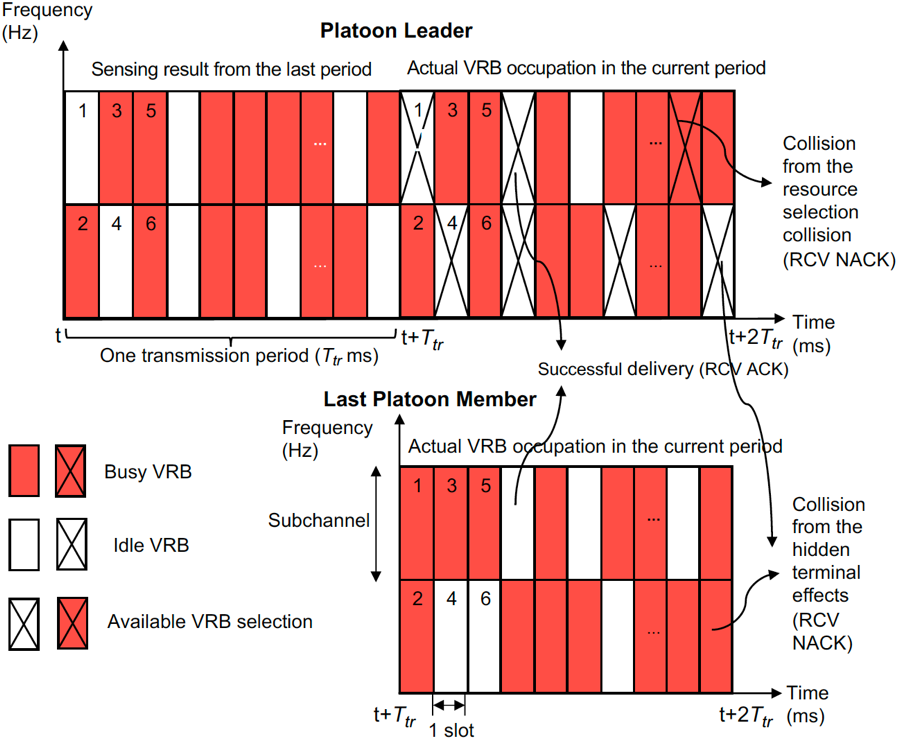}
    \caption{The VRB maps of resource allocation for platoon communication in one transmission period: the available VRB selections in the current transmission period correspond to the idle VRBs sensed from the last transmission period. PL only selects one of those VRBs even some of them are occupied by other vehicles due to the resource selection collision or the hidden terminal effects in the current transmission period. The VRB in each transmission period is indexed from 1 to $N_r$ as above.}
    \label{fig:RB_selection}
\end{figure}

As shown in Fig. \ref{fig:RB_selection}, in transmission period $n$ (i.e., $n = 1,2,...$ ), PL selects an action $a_{n} \in \mathcal{A}_{n-1} \subseteq \mathcal{A}$, where $\mathcal{A}$ is the action set of all VRBs in one transmission period while $\mathcal{A}_{n-1} $ is the action set of the VRBs sensed idle in transmission period $n-1$. In other words, if we denote the fixed set $\mathcal{A} = \{ VRB_1, VRB_2, ..., VRB_{N_r}\}$, where each VRB is indexed, and $\mathds{1}_{idle}(m)$ indicates whether $VRB_m$ is sensed idle. Thus $\mathcal{A}_{n-1} $ can be described as
\begin{equation}\label{eq:A_{n-1}}
    \mathcal{A}_{n-1} = \{VRB_m |  VRB_m \in \mathcal{A} \  and \ \mathds{1}_{idle}(m)=1\}.
\end{equation}
 Note that $\mathcal{A}_{n-1}$ is a dynamic set with transmission period $n-1$ varying due to the resource keeping probability in SPS scheme. Then PL observes the received feedback, $o_n \in \{ACK,NACK\}$ before the end of this period, and utilizes the historical observations and its own past actions, to select the next action $a_{n+1} \in \mathcal{A}_{n}$ in transmission period $n+1$. 
 
 Next, we denote the action-observation tuple of PL in transmission period $n$ as $c_{n} = \{ a_{n}, o_{n} \}$. Denote the environment history of PL as the state in transmission period $n$, $s_{n} = [c_{n-M+1}, ... ,c_{n-1}, c_{n} ]$ which is a combination of historical selected idle VRBs as well as its observed feedback over previous $M$ transmission periods. Based on $s_n$, PL takes an action $a_{n}$ and the state $s_n$ transfers to $s_{n+1}$ with reward $r_{n+1}$. The reward $r_{n+1}$ is determined by the observed feedback: If an ACK is received, PL will be given by a positive reward ($r_{n+1} > 0$) while if a NACK is received, PL will be rewarded nothing ($r_{n+1} = 0$). 
As illustrated above, the state $s_{n+1}$ only depends on $s_{n}$ after PL takes an action $a_{n+1}$, the state description of PL can be used to construct a Markov decision process (MDP) formulation.

\subsection{Deep Reinforcement Learning Framework}
In deep reinforcement learning (DRL) algorithm, the goal of PL is to seek a policy that maximizes the expected long-term number of successful transmissions. The cumulative reward of PL in transmission period $n$ can be expressed as 
\begin{equation}
    R_n=\sum_{k=0}^{\infty} \gamma^{k} r_{n+k},
\end{equation}
where $\gamma \in \left(0,1\right]$ is a discounting factor. Now, we introduce the action-value function of PL, $Q^{\pi}(s, a)$, which is defined as the expected sum of the discounted reward received starting from the state $s$, taking the action $a$ and then following the policy $\pi$ thereafter, which can be formally expressed as
\begin{equation}
    Q^{\pi}(s, a)=\mathbf{E}[R_n|s_n=s ,a_n=a; \pi],
\end{equation}
If we denote $Q^{\pi^*}(s,a)$ as the action-value with initial state $s$ and action $a$ and then following the optimal policy $\pi^*$, the optimal policy $\pi^*$ can be derived as
\begin{equation}
\begin{aligned}
    \pi^*(s) = \text{arg max}_aQ^{\pi^*}(s,a), \forall s.
\end{aligned}
\end{equation}
$Q^{\pi^*}(s,a)$ can be obtained by letting PL update $Q(s_n,a_n)$ in each transmission period. The update rule of $Q(s_n,a_n)$ is given as follows:
\begin{equation}
    \begin{aligned}
        Q(s_n,a_n) &\leftarrow  Q(s_n,a_n)\\ &+\alpha \left[r_{n+1}+\gamma \max_{a_{n+1}}Q(s_{n+1},a_{n+1})-Q(s_{n},a_{n})\right],
    \end{aligned}
\end{equation}
where $\alpha \in \left(0,1\right]$ denotes the learning rate. As introduced beforehand, in transmission period $n$, the action $a_n$ transfers the state $s_n$ to $s_{n+1}$ which gives the reward $r_{n}$. At the end of this transmission period, PL can also obtain the information of $\mathcal{A}_{n}$  which is the set of available actions for the next period. The above information, $(s_{n},a_{n},r_{n},s_{n+1},\mathcal{A}_{n})$, hence forms a single batch of training set for QNN. Given the reward information $r_n$, Q-learning Neural Network (QNN) can be trained by minimizing the prediction loss function $L(\text{\boldmath$\theta$})$, where $\text{\boldmath$\theta$}$ denotes the parameters to be trained of QNN. The prediction loss function is given by 
\begin{equation}
    L(\text{\boldmath$\theta$})=\mathbf{E}\left\{\left[v(r_{n},s_{n+1}, \mathcal{A}_{n})-Q(s_{n},a_{n};\text{\boldmath$\theta$})\right]^2\right\},
\end{equation}
where $Q(s_{n},a_{n};\text{\boldmath$\theta$})$, as the predicted value function, is the output of QNN, meanwhile, the target value function is defined as 
\begin{equation}
    v(r_{n},s_{n+1},\mathcal{A}_{n})=r_{n}+\gamma \max_{a_{n+1}\in \mathcal{A}_{n}} Q(s_{n+1},a_{n+1};\text{\boldmath$\theta$}),
    \label{eq:gam}
\end{equation}
where the second term $\max Q(s_{n+1},a_{n+1};\text{\boldmath$\theta$})$ is obtained by searching the maximum output of QNN with respect to the selection of available action $a_{n+1}$ given $s_{n+1}$. 

While PL updates $Q(s_n,a_n)$, it also allocates the resource for its communication with PMs based on $Q(s_n,a_n)$ in real time. With the $\epsilon$-greedy policy, PL selects an action with the highest action value from $\mathcal{A}_{n}$ with probability $1-\epsilon$, otherwise, it randomly selects an action from $\mathcal{A}_{n}$. Thus, the optimal action $a_{n+1}^*$ is expressed as 
\begin{equation}
        a_{n+1}^*=
        \begin{cases}
         \text{arg max}_{a_{n+1}\in \mathcal{A}_{n}}Q(s_{n+1},a_{n+1}), &P=1-\epsilon.\\
         a_{n+1} \in \mathcal{A}_{n}, &P= \epsilon,
        \end{cases}
\label{eq:eps}
\end{equation}
where $\epsilon$ is the probability of selecting an action randomly. The $\epsilon$-greedy policy applied in the DRL algorithm has been shown to be effective in avoiding saddle points by trying new random actions \cite{sutton2018reinforcement}.

\section{Simulation}
\label{sec:sim}
This section entails the analytical results and simulations in terms of the vehicle density and the resource keeping probability for the random selection algorithm and our proposed DRL algorithm. We first use Matlab to conduct Monte Carlo simulation for all broadcast vehicles adopting the SPS scheme on a finite length road with 1-dimension based on a simplified version of Mode 2. Concurrently, we generate the data set that is utilized in the simulation of resource allocation for platoon communication. Subsequently, we test our analytical model to validate our simulation architecture which is based on the generated data set. Finally, we develop the DRL algorithm in the same simulation architecture and compare the collision probability on the same data set with the random selection algorithm.

On the generated data, we utilize Python to implement the random selection algorithm and the DRL algorithm with PyTorch and then estimate $P_{c}^{ht}$ under the different cases. The simulations were implemented on a server with a CPU (Intel Core i9-9700k) and a GPU (NVIDIA GeForce GTX 2080Ti) \footnote{The simulation code can be accessed at https://github.com/Mauriyin/V2V-Groupcast}. The detailed simulation parameters and the hyper-parameters for the deep neural network we adopted for the simulation are introduced in the following sections.
 
\begin{table}[]
\caption{Simulation Parameters}
\label{tb:simu_pra}
\centering
\resizebox{.4\textwidth}{!}{
\begin{tabular}{|c|c|}
\hline
\textbf{Parameters}         & \textbf{Value}  \\ \hline
Center frequency $f_c$ (GHz) & 30 \\ \hline
Sub-carrier spacing (kHz) & 30 \\ \hline
Transmit power $P_{trans}$ (dBm) & 23 \\ \hline
Sensitivity power level $P_{sen}$ (dBm) & -91 \\ \hline
Highway length (km) & 4 \\ \hline
Vehicle density $\rho$ (vehicles/km) & \{20,40,...,200\} \\ \hline
Platoon length $d$ (km) & 0.1 \\ \hline
Transmission periodicity $T_{tr}$ (ms)  & 50   \\ \hline
Number of sub-channels $n_r$ & 2 \\ \hline
Duration of one slot $t_s$ (ms) & 0.5 \\ \hline
\tabincell{c} {Number of transmission periods \\ in one SPS period $T_s$} & \tabincell{c} {10} \\ \hline
Resource keeping probability $p$ & \{0.9, 0.7, 0.5\} \\ \hline
\tabincell{c} {Length of one simulation run \\ (Transmission periods)} & \tabincell{c} {10000} \\ \hline

\end{tabular}
}
\end{table}

\subsection{Data Generation}
Table I articulates the simulation parameters. In each transmission period, we only capture the sensing results of the PL and the last PM. The sensing result of the PL or the last PM in one transmission period is just one row of 0s and 1s indicating the idle VRBs and busy VRBs, indexed from left to right. For each experiment, the generated data from the PL and the last PM are the matrices in $\mathbf{R}^{10000 \times 200}$. Since NR-V2X can support the system bandwidth up to 100 MHz, each VRB in one slot can accommodate any periodic BSM delivery in the simulation. In each experiment, the location of each vehicle is determined by a uniformly distributed random variable with the maximum value being the length of road. In addition, we set the vehicle near the middle of lane as the PL in order to avoid the edge effect of hidden terminals on the performance of platoon communication. We conduct 50 different experiments in each pair of vehicle density and resource keeping probability to average the relevant results.

\subsection{Random Selection}
\begin{figure}[tp]
    \centering
    \includegraphics[width=.45\textwidth]{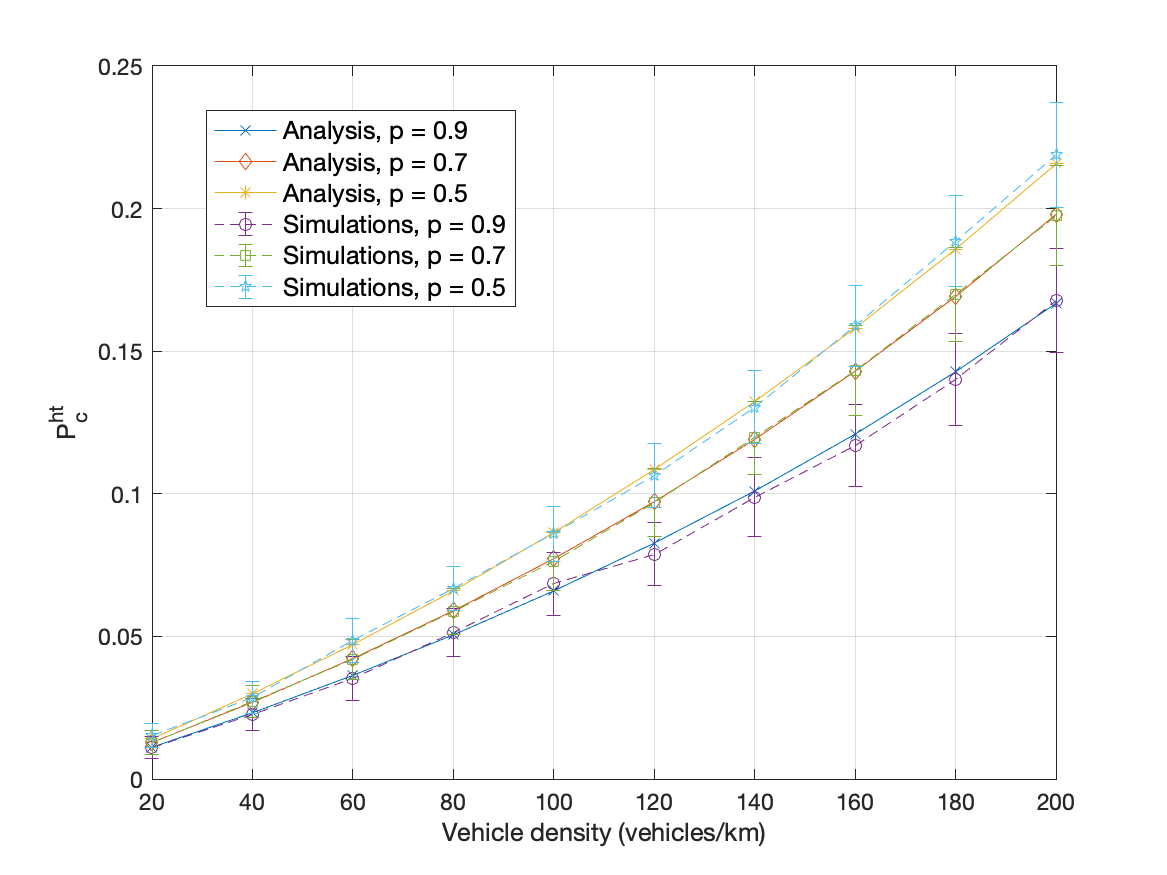}
    \caption{Comparison between analytical and simulation results of $P_c^{ht}$ for random selection algorithm.}
    \label{fig:result_1}
\end{figure}

We first test the analytical model proposed in Section \ref{sec:sys}. The simulation results under the different vehicle density and resource keeping probability are illustrated in Fig. \ref{fig:result_1}. It is clear that the simulation results are consistent with analytical results in different parameters. When the resource keeping probability is fixed, if the vehicle density increases, $P_c^{ht}$ also increases due to the increasing number of hidden terminals. Meanwhile, when the vehicle density is fixed, if the resource keeping probability decreases, the vehicles are more likely to change their VRB selections and select the same VRB as PL in a transmission period, leading to a higher $P_c^{ht}$. From an analytical perspective, $P_c^{rs}$ expressed in equation (\ref{eq:P_c^sr}) will increase when the resource keeping probability $p$ decreases, which thereby increases $P_c^{ht}$ expressed in equation (\ref{eq:P_c^ht}). Since the random selection algorithm only takes advantage of the sensing results of PL, the increment of $P_c^{ht}$ seems to be rapid when the vehicle density increases. However, with the feedback from PMs, PL can learn from the environment that helps PL lower $P_c^{ht}$.

\subsection{Deep Reinforcement Learning}
We develop the DRL algorithm introduced in Section \ref{sec:DRL} within the same simulation architecture and data set employing the random selection algorithm. In each transmission period, PL collects the historical actions, i.e., the selected VRBs, and the historical observations, i.e., the received feedback (ACK/NACK), over previous 16 transmission periods to construct the state, and then selects the VRB with the highest Q-value provided from the deep neural network among the idle VRBs sensed from the last transmission period.

\begin{figure}[htp]
    \centering
    \includegraphics[width=.18\textwidth]{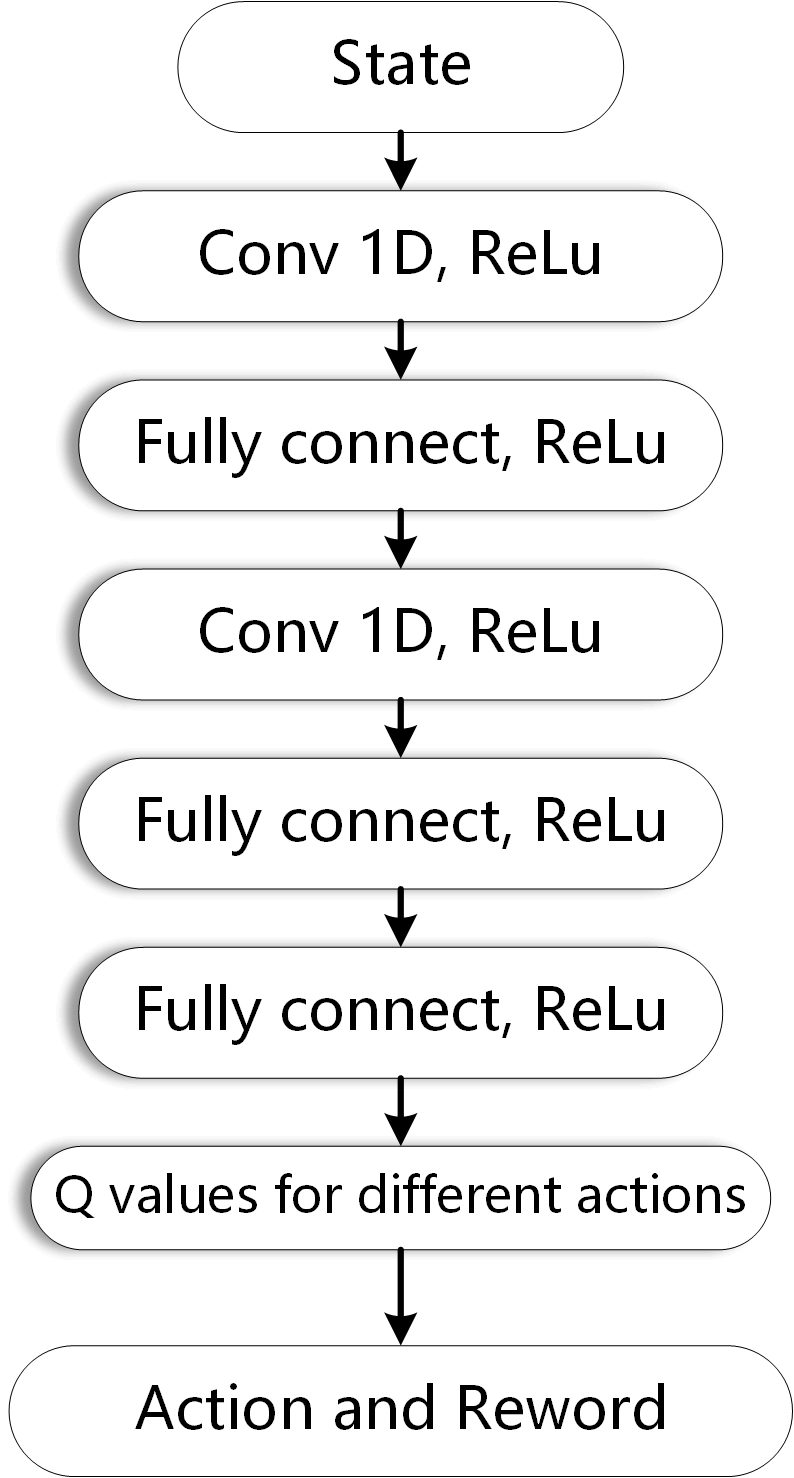}
    \caption{Deep Q-learning Neural Network. \\
    The architecture of QNN is a five-layer deep QNN.}
    \label{fig:dqn}
\end{figure}

We construct a five-layer deep QNN, with two 1-dimensional convolution layers and three fully connected layers, as illustrated in Fig \ref{fig:dqn}. For the system model detailed in Section \ref{sec:DRL}, the state dimension and input data are relatively small in the examples simulated, we thus choose a five-layer deep QNN which is not too deep so that the model weight can be controlled. Table \ref{tb:hyper_pra} lists the hyper parameters of the deep Q-learning neural network. Note that the simulation is running from the very beginning, and we do not pre-train the DRL model on the same data but let the model train during the process. The collision is calculated from the same start transmission period as the random selection algorithm so that the comparison between the two discussed algorithms is fair (no extra training or learning time for DRL). 

\begin{table}[]
\caption{Hyper-Parameters of QNN}
\label{tb:hyper_pra}
\centering
\resizebox{.22\textwidth}{!}{
\begin{tabular}{|c|c|}
\hline
\textbf{Parameters}         & \textbf{Value}  \\ \hline
State size & 32 \\ \hline
Batch size & 1 \\ \hline
Action size & 200 \\ \hline
Learning rate & 0.01   \\ \hline
$\gamma$ in eq (\ref{eq:gam}) & 0.9 \\ \hline
$\epsilon$ in eq (\ref{eq:eps}) &  1 \\ \hline
$\epsilon$ minimum value & 0.0 \\ \hline
$\epsilon$ decay & 0.5 \\ \hline
Memory size & 1000 \\ \hline

\end{tabular}
}
\end{table}

\begin{figure}[htp]
    \centering
    \includegraphics[width=.45\textwidth]{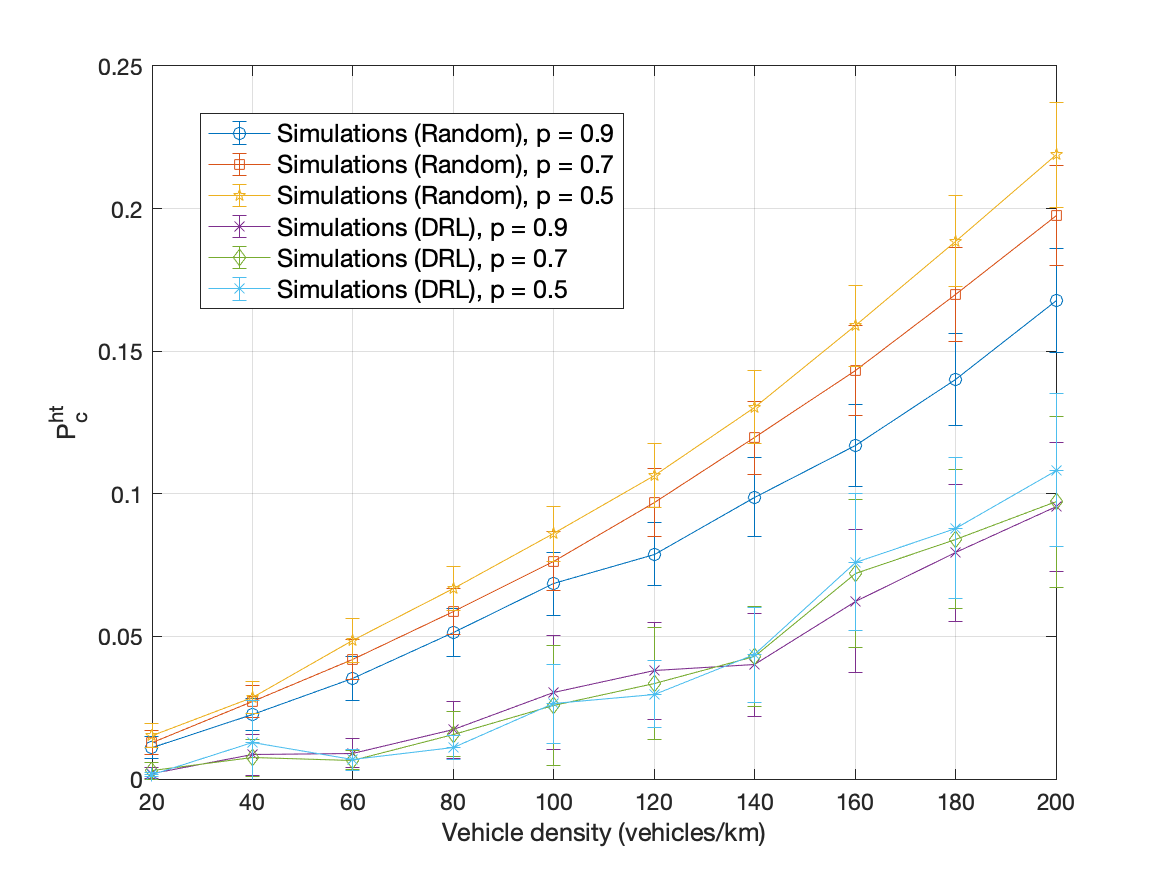}
    \caption{Simulation comparison between random selection and DRL algorithm.}
    \label{fig:result_2}
\end{figure}
The simulation results are shown in Fig. \ref{fig:result_2}. With the same resource keeping probability, $P_c^{ht}$ applied by DRL is always lower than that with the random selection algorithm in different vehicle density. In particular, $P_c^{ht}$ in the low and high vehicle density is at least decreased by 73\% and 45\% respectively. When the vehicle density increases, $P_c^{ht}$ also increases, nevertheless, it's still lower than that applied by the random selection algorithm significantly. Moreover, DRL seems to be less sensitive regarding the change of resource keeping probability. Since DRL maintains a memory size of each state, training the neural network from the memory, PL can thereby learn the potential pattern of the communication environment, which is less likely to be impacted by some changes such as the different resource keeping probability in the environment.

% Therefore, DRL always outperforms the random selection algorithm by constructing the state including the historical actions and observations from the communication environment.     

\begin{figure}[tp]
    \centering
    \includegraphics[width=.45\textwidth]{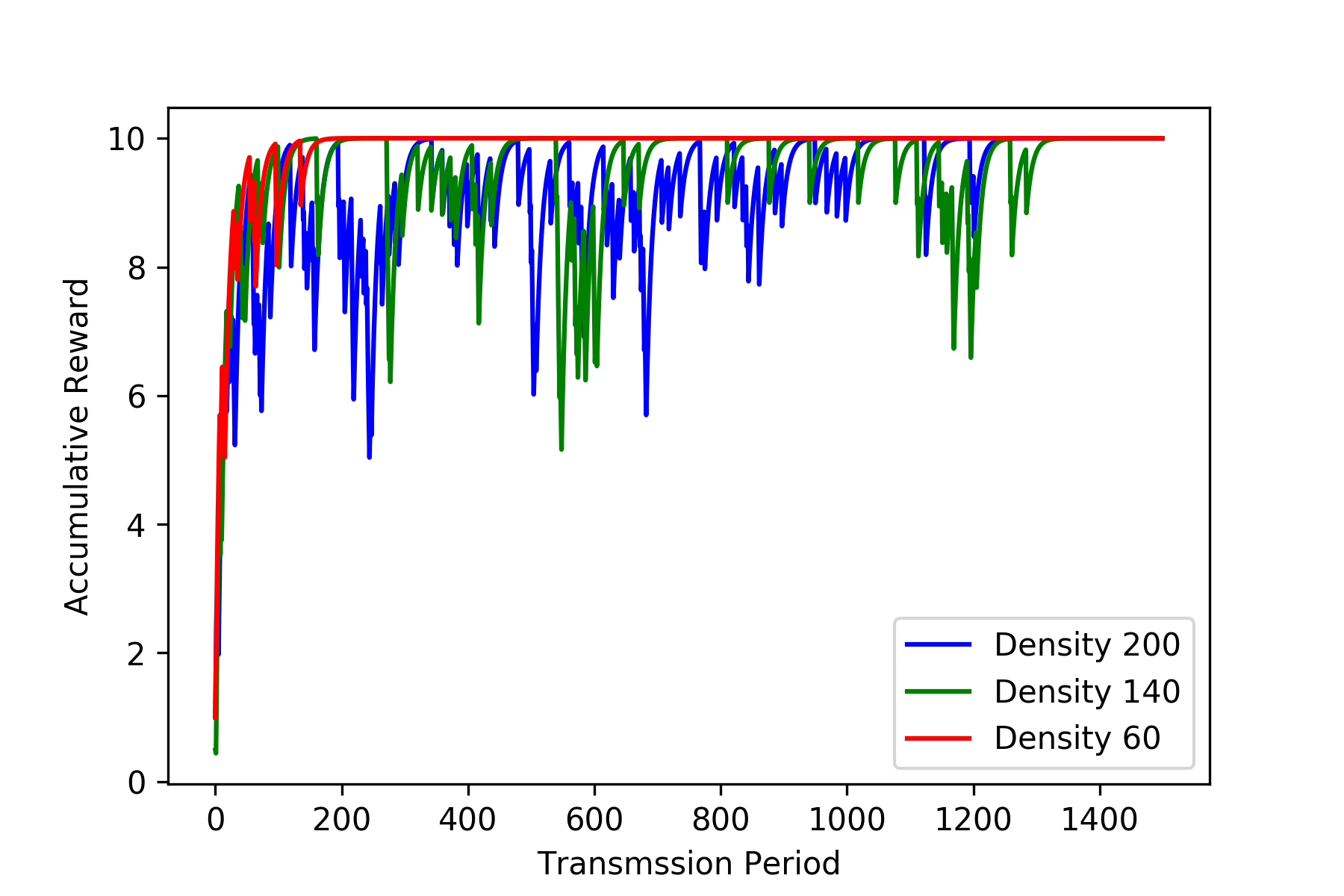}
    \caption{Cumulative reward as time increasing for different vehicle density with keeping probability $p=0.5$.}
    \label{fig:reward}
\end{figure}

The convergence speed of reward with the keeping probability $p = 0.5$ in different vehicle density is shown in Fig. \ref{fig:reward}. The coverage speed for each vehicle density is around 200 transmission periods. As the vehicle density increases, it is harder to capture the hidden terminal effects that the performance of DRL becomes less stable. The impact of the learning period is negligible compared with the length of the simulation time.

\section{Conclusion}
\label{sec:con}
In this paper, we focused on the resource allocation for vehicle platooning from the MAC-layer perspective in the out-of-coverage scenario where multiple communication types coexist, featuring the new characteristics of NR-V2X. We analyzed the collision probability of platoon communication via the random selection algorithm. We then applied the DRL algorithm to decrease the collision probability by letting PL learn from the communication environment. In addition, we utilized Monte Carlo simulation to verify the analytical model and compare the results between the two discussed algorithms. The results indicate that our analytical model provisions a good estimate for the collision probability of the random selection algorithm. Meanwhile, our proposed DRL algorithm outperforms the random selection algorithm by decreasing the collision probability in terms of different vehicle density and resource keeping probability.

In our future work, it is possible to explore more results in the 2-dimensional scenario in which the width of the multi-lane highway cannot be negligible. In such a scenario, different PM in the platoon experiences different hidden terminal effects, and all of them will impact $P_c^{ht}$ simultaneously. 

\bibliographystyle{IEEEtran}
\bibliography{ref}
\end{document}